\documentclass[12pt]{iopart}
\usepackage{graphicx}
\usepackage{rotating}
\usepackage{amsfonts}
\usepackage{amssymb}
\usepackage{enumerate}
\usepackage{longtable}
\usepackage{dcolumn}

\usepackage{bm}

\begin{document}

\title{Preformed Excitons, Orbital Selectivity, and Charge-Density-Wave 
Order in 1T-TiSe$_{2}$}

\author{S. Koley$^{1}$, M. S. Laad$^{3}$, N. S. Vidhyadhiraja$^{4}$ and A. 
Taraphder$^{1,2}$}
\address{$^{1}$Department of Physics and Centre for Theoretical Studies,\\ 
Indian
Institute of Technology, Kharagpur 721302 India\\
$^{2}$ School of Basic Sciences, Indian Institute of Technology, Mandi 175001 
India \\
$^{3}$Institut Laue Langevin, 6, Rue Jules Horowitz, 38042 Grenoble Cedex, France. 
\\
$^{4}$ Jawaharlal Nehru Centre for Advanced Scientific Research, Bangalore
560064 India}

\begin{abstract}
Traditional routes to Charge-Density-Wave in transition metal dichalcogenides, relying on 
Fermi surface nesting or Jahn-Teller instabilities have recently been
brought into question. While this calls for exploration of alternative 
views, paucity of theoretical guidance sustains lively controversy on the
origin of, and interplay between CDW and superconductive orders in 
transition metal dichalcogenides. Here, we explore a preformed excitonic liquid 
route, heavily supplemented by modern correlated electronic structure
calculations, to an excitonic-CDW order in 1T-TiSe$_{2}$. We show that 
orbital-selective dynamical localisation arising from preformed excitonic liquid
correlations is somewhat reminiscent to states proposed for $d$- and $f$-band quantum
criticality at the border of magnetism. Excellent quantitative
explication of a wide range of spectral and transport
responses in both normal and CDW phases provides strong support
for our scenario, and suggests that soft excitonic liquid 
fluctuations mediate superconductivity in a broad class of 
transition metal dichalcogenides on the border of CDW. This brings the 
transition metal dichalcogenides closer
to the bad actors in $d$- and $f$-band systems, where anomalously
soft fluctuations of electronic origin are believed to mediate 
unconventional superconductivity on the border of magnetism.
\end{abstract}
\maketitle

\section{Introduction}
Twenty six years after the path-breaking discovery of high-$T_{c}$
superconductivity in doped, quasi-two dimensional
copper oxides (cuprates), the list of ill-understood strongly
correlated electronic systems (SCES) with partially filled $d$ or $f$
bands continues to grow~\cite{htsc,steg,imada}.  The paradigm
shift engendered by cuprates also fuelled renewed interest and study of 
older systems. Before the cuprate revolution, these were
sore spots in the standard model of electrons in metals, centred
around the celebrated Landau Fermi liquid
(LFL) theory. Intense activity to unravel increasing variety of
unconventional ordered states with ill-understood
metallicity has stabilised this paradigm shift. Cuprates are not, it
is now clear, an isolated example: rare-earth
systems close to magnetic instabilities at $T=0$, increasing number of
$d$-band perovskites and the recent
explosion in Fe-arsenides are but a few examples of a truly diverse
zoo of strange systems.

Perhaps equally remarkable is the fact that careful work in the
recent past has brought out unexpected similarities between the newer and 
older bad actors above. This is best exemplified by the recent revival of 
interest in quasi-2D transition metal dichalcogenides (TMD).
For more than forty years, the origin of charge density wave (CDW) and superconductive (SC) 
orders and their interplay in the layered TMD had remained ill-understood
issues~\cite{wilson,rice}.  Historically, appealing to one-electron
(density-functional theory) band structure led to the conventional wisdom of 
CDW arising from Fermi surface (FS) nesting, or via a band Jahn-Teller (JT)
instability. Recent revival in the field was stimulated, among other
things, by falsification of these views by recent ARPES work~\cite{monney1}. 
The exciting possibility of an alternative, intrinsically strong coupling 
view, involving CDW order emerging as a Bose-Einstein condensation (BEC) of
an incoherent preformed excitonic liquid (PEL) normal state, 
arose~\cite{arghya,monney2,wezel} as an attempt to address this new conflict,
and has the potential to bring TMD into the list of strange actors. 
Indeed, parallels between TMD and cuprates have increasingly been claimed in
certain studies~\cite{bsenko}. It is also suggested that the absence
of magnetism in TMD allows studies of a bad metal without attendant magnetic 
fluctuations that complicate the physics in cuprates~\cite{dordevich}.   
That a strong coupling scenario for TMD closer to the Mott limit is in 
order, is shown by the
fact that the related system 1T-TaS$_{2}$ has long been understood as
a Mott insulator~\cite{tosatti}: it is then fully conceivable that
Mottness also plays important roles in other TMD.  From this
viewpoint, much as in the more recent examples of intense interest,
melting of the excitonic-CDW order in a PEL scenario would necessarily
enhance excitonic fluctuations, which could act as an
unconventional electronic `pairing glue' for emergence of the
competing SC order. However, the theoretical situation still remains
unclear, and FS-nesting as well as strong electron-phonon views are still
claimed to be the instigators of CDW order and
SC in pressurised or doped systems\cite{bsenko}.

The long-studied 1T-TiSe$_{2}$ is a particularly relevant case in
point. Careful experimental work has unravelled behavior that 
consistently fails to fit into any conventional views. We list the 
problems here:\\
(i) While optical
studies~\cite{opt} show that the normal-CDW
transition is a semi-metal to semiconductor transition a la the
Overhauser scenario, noticeable spectral weight
transfer (SWT) over an energy scale of order 1.0$~eV$ upon heating from 10K
to 300K, along with a clear isosbestic point as a function of temperature ($T$) below 
$T_{cdw}=200$~K and a large normal state scattering rate support
sizable electronic correlations.\\
(ii) Pure 1T-TiSe$_{2}$ shows
bad-metallic resistivity, much above the Mott limit,
even at pretty low $T ~ (<<T_{cdw}$), with a maximum, but no anomaly, in
$d\rho/dT$ at $T\simeq T_{cdw}$~\cite{sipos}. Tellingly, recovery of
good metallicity at high pressures also destroys SC. Such a
correlation is repeatedly seen in many systems where unconventional SC 
appears near Mott insulators and magnetism~\cite{hussey} (however, 
TMD never show magnetism, nor has Mottness been hitherto considered 
important, except for 1T-TaS$_{2}$~\cite{tosatti} and 2H-TaSe$_2$\cite{arghya,dordevich}). 
Thus, whatever 
causes strong normal state scattering also facilitates SC pair formation 
below $T_{c}$.\\ 
(iii) ARPES data~\cite{aebi} shows that the
renormalised electronic structure in the normal state
($T>T_{cdw}$) already resembles an LDA dispersion modified by 
excitonic correlations, supporting a PEL view. Moreover,
lack of clear polaron effects (e.g, kinks in the band dispersion at 
phonon energies, well below E$_F$ in resonant inelastic
X-ray scatteing~\cite{monney2}) and, more importantly, the
insensitivity of ARPES band dispersions to
atomic displacements occuring across $T_{cdw}$, argue against a
band-JT instability, even though electron-lattice
coupling is ubiquitous to TMD~\cite{rolf}.\\   
(iv) Finally, in spite of the
incoherent metal features pointed out above, the effective mass is
only weakly renormalised above $T_{cdw}$.

These observations put strong constraints on an acceptable theory.
Most importantly, they conflict with both FS
nesting and band-Jahn-Teller (JT) views on general grounds:  Absence
of band quasiparticles in the normal state
rules out a conventional ordering instability involving the LDA Fermi
surface (FS) features.  Simply put, the very
concept of a well-defined FS in the LDA sense becomes untenable in
bad metals.  As pointed out above, ARPES data also argue against a
band-JT instability to CDW order. Weak mass renormalisation 
above $T_{cdw}$ conflicts with one-band modeling, but not with a
multi-band approach~\cite{held}: even in the classic Mott case of
$V_{2}O_{3}$, the effective mass in the correlated metal is 
only moderately enhanced near the (undoubtedly correlation driven) 
metal-insulator transition~\cite{held,laad}. Thus,
taken together, these observations force one to view emergence of the
CDW as a strong coupling Overhauser instability of a multi-band 
bad-metal normal state without LFL quasiparticles.

These issues have been selectively addressed earlier within conflicting
theoretical views, but the above observations force us to
critically re-examine them. Here, we show that these anomalous responses 
are naturally understood within our PEL view~\cite{arghya} for 
1T-TiSe$_{2}$, strongly supporting the PEL as a novel (and perhaps generic) 
alternative to conventional theories for TMD. In remarkable parallels 
with $d$ and $f$-band systems close to Mottness and unconventional order(s), 
we unravel a new instance of a selective metallic state in the PEL, 
and construct a scenario for the instability of this PEL to a 
low-$T$ CDW state. Our philosophy is opposite in spirit to weak-coupling 
Fermi liquid ones, and is actually closer to resonating valence bond (RVB) 
ideas~\cite{phil-book}, in the sense that ordered state(s) arise as 
two-particle (BEC) instabilities of an incoherent liquid of preformed 
excitons.

\section{Method}
\begin{figure} 
\centering
{\includegraphics[angle=270,width=0.8\columnwidth]{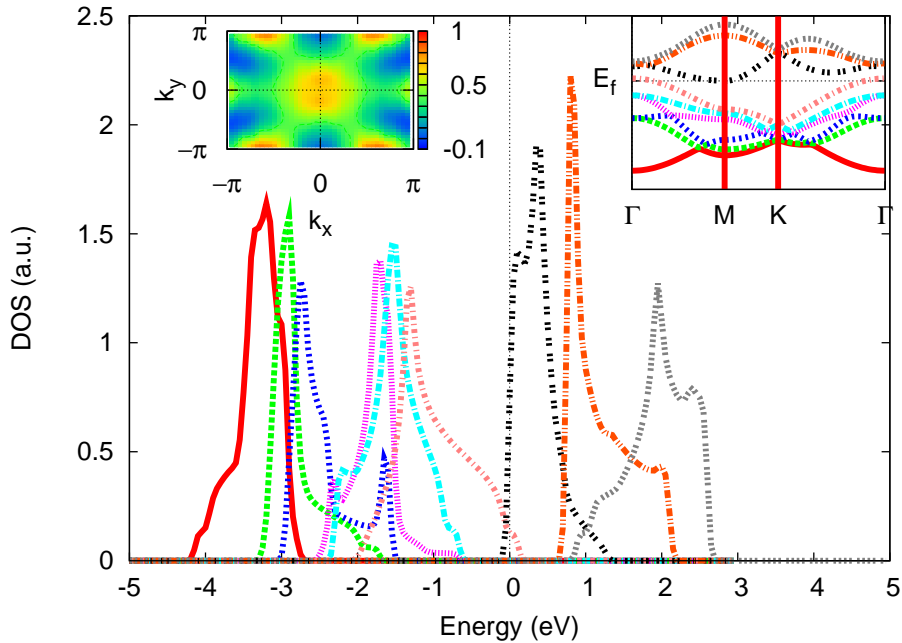}}
\caption{(Color Online) Non interacting DOS, tight binding band structure for 
the Ti-d $t_{2g}$ and Se-p bands and the corresponding Fermi surface plot. 
The MO DMFT involves Ti-d (black) and Se-p (yellow) bands.}
\label{fig1}
\end{figure}

LCAO band structure for 1T-TiSe$_{2}$ was constructed~\cite{wezel,yoshida}
by using the Ti-$d$ and Se-$p$ states. This gives two bands closest 
to $E_{F}$ (predominantly Ti-$d_{xy}$ and Se-$p_{z}$) as well as 
the FS in very good accord with earlier results~\cite{yoshida}, as shown 
in Fig.~\ref{fig1}. A sizable $d_{xy}-p_{z}$ mixing hybridises the
small number of electrons and holes.  In this situation, even moderate
electronic correlations ($<1.0$~eV) facilitate exciton formation 
already at high $T$, as already discussed by Halperin {\it et al.}~\cite{rice}
in the sixties. Given that electronically active states comprise $d_{xy}$
and $p_{z}$ band states, an intraband Hubbard $U\simeq 1.0$~eV and interband 
correlation $U_{ab}\simeq 0.5-0.7$~eV are realistic values: these can 
be estimated from a
first-principles, constrained-LDA calculation, and we believe~\cite{yaresko} 
that these fall in the range quoted above. 

Though static 
Hartree-Fock (HF) theory can now give a CDW {\it ground}
state~\cite{aebi,wezel}, description of observed normal state
incoherence, and in general, preformed liquid-like electronic
states, lies outside its scope.  Thus, the character and ordering
instabilities of such bad metals {\it cannot}, by construction, 
be rationalised by appealing to static-mean-field theory: this can 
only be reliably accessed by approaches which can adequately 
capture dynamical correlations. For 1T-TiSe$_{2}$, a minimal 
two-band Hubbard model as defined below is mandated by LCAO results, and 
adequate treatment of dynamical correlations underlying incoherent behavior 
is achieved by dynamical-mean field theory (DMFT). DMFT and cluster-DMFT 
approaches have, by now, a proven record of successfully treating 
strong dynamical fluctuatons in correlated electronic systems, making them
preferred tools of choice in the present context. The two-band Hubbard model 
we use is (calling $d_{xy}=a$ and $p_{z}=b$)

$$H_{el}=\sum_{{\bf k},l,m,\sigma}(t_{\bf
k}^{lm}+\epsilon_{l}\delta_{lm})c_{{\bf k}l\sigma}
^{\dag}c_{{\bf k}m\sigma}+
U\sum_{i,l=a,b}n_{il\uparrow}n_{il\downarrow}
+ U_{ab}\sum_{i}n_{ia}n_{ib}$$\\
\noindent where $l,m$ run over both band indices $a,b,$ the intra-orbital correlation is $U$ 
(taken to be same for $a,\, b$ bands, we have checked that results are insensitive
to this choice within reasonable limits), $U_{ab}$ is the
inter-orbital correlation term that, along with
$t_{\bf k}^{ab}$, will play a major role throughout. Further, in TMD, 
the most relevant $A_{1g}$ phonon mode couples to the inter-band
excitons~\cite{wezel} by symmetry, and the electron-phonon coupling is
$H_{el-l}=g\sum_{i}(A_{i}+A_{i}^{\dag})(c_{ia}^{\dag}c_{ib}+h.c)$. 
To solve $H=H_{el}+H_{el-l}$ within DMFT, we have combined the multi-orbital 
iterated perturbation theory (MOIPT) for $H_{el}$~\cite{laad} with the DMFT 
for polarons by Ciuchi {\it et al.}~\cite{ciuchi,aipp} (see Appendix). 
Actually, this involves extending the polaron-DMFT to multiband cases, seen 
by writing $H_{el-l}=g\sum_{i}(A_{i}+A_{i}^{\dag})(n_{i,+}-n_{i,-})$ by a
rotation, $c_{i,\pm}=(c_{i,a}\pm c_{i,b})/\sqrt{2}$ and $n_{i,\pm}=c_{i,\pm}^{\dagger}
c_{i,\pm}$. Finally, as a theoretical advance, we extend the normal state DMFT 
to the broken-symmetry CDW phase as before~\cite{arghya}. This is justified, since, from 
the above discussion and LCAO+DMFT results below, we find, {\it a posteriori}, an incoherent 
PEL. Instability to CDW order then cannot occur via the traditional band-folding 
of well-defined Fermi liquid (FL) quasiparticles. Rather, as coherent 
one-particle inter-band mixing is inoperative, ordered states must now arise
directly as two-particle instabilities of the bad metal. For 1T-TiSe$_{2}$, 
the residual two-particle interaction, obtained to second order is 
proportional to $t_{ab}^{2}$, more relevant than the (incoherent)
one-electron mixing, $t_{ab}$. The interaction  
$H_{res}\simeq -t_{ab}^{2}\chi_{ab}(0,0)\sum_{<i,j>,\sigma\sigma'}
c_{ia\sigma}^{\dag}c_{jb\sigma}c_{jb\sigma'}^{\dag}c_{ia\sigma}$,
with $\chi_{ab}$ the inter-orbital susceptibility. Decoupling
this intersite interaction in a generalised HF sense yields two 
competing instabilities: $H_{res}^{(HF)}=-\sum_{<i,j>,
\sigma\sigma'}(\Delta_{1b}c_{ia\sigma}^{\dag}c_{ia\sigma}
+ \Delta_{ab}c_{ia\sigma}^{\dag}c_{jb-\sigma}^{\dag} + a\rightarrow
b)$, with $\Delta_{cdw}=(\Delta_{1a}-\Delta_{1b})\propto \langle n_{a}-n_{b}
\rangle$ representing a CDW and $\Delta_{ab}\propto \langle c_{ia\sigma}
c_{jb-\sigma}\rangle$ a multiband spin-singlet SC. Following earlier
procedure~\cite{arghya}, we compute DMFT spectral functions and transport 
properties in the CDW state at low $T$, leaving SC for future. 

\section{Results}
\begin{figure}
\centering
(a)
{\includegraphics[angle=270,width=0.6\columnwidth]{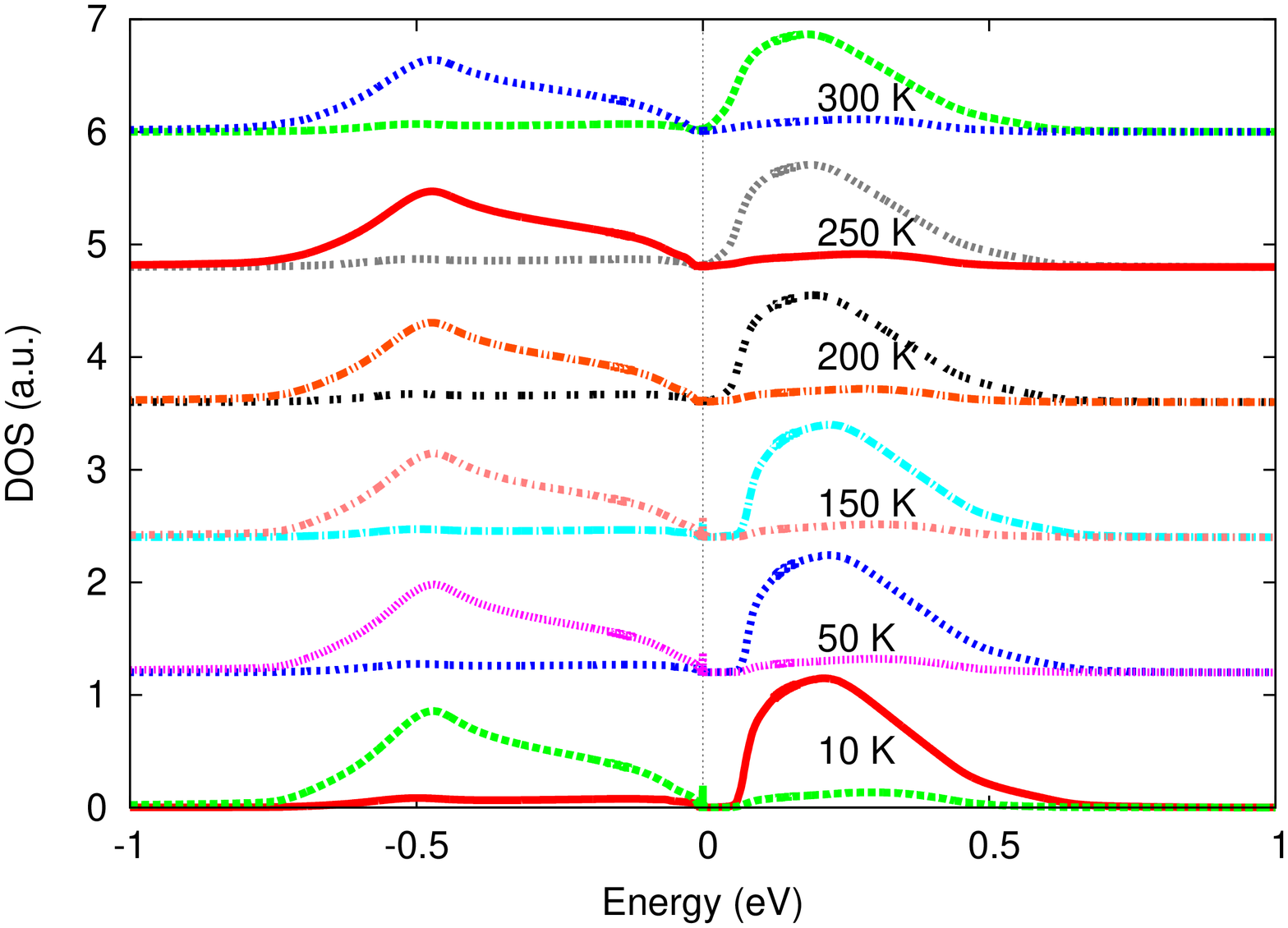}}

(b)
{\includegraphics[angle=270,width=0.6\columnwidth]{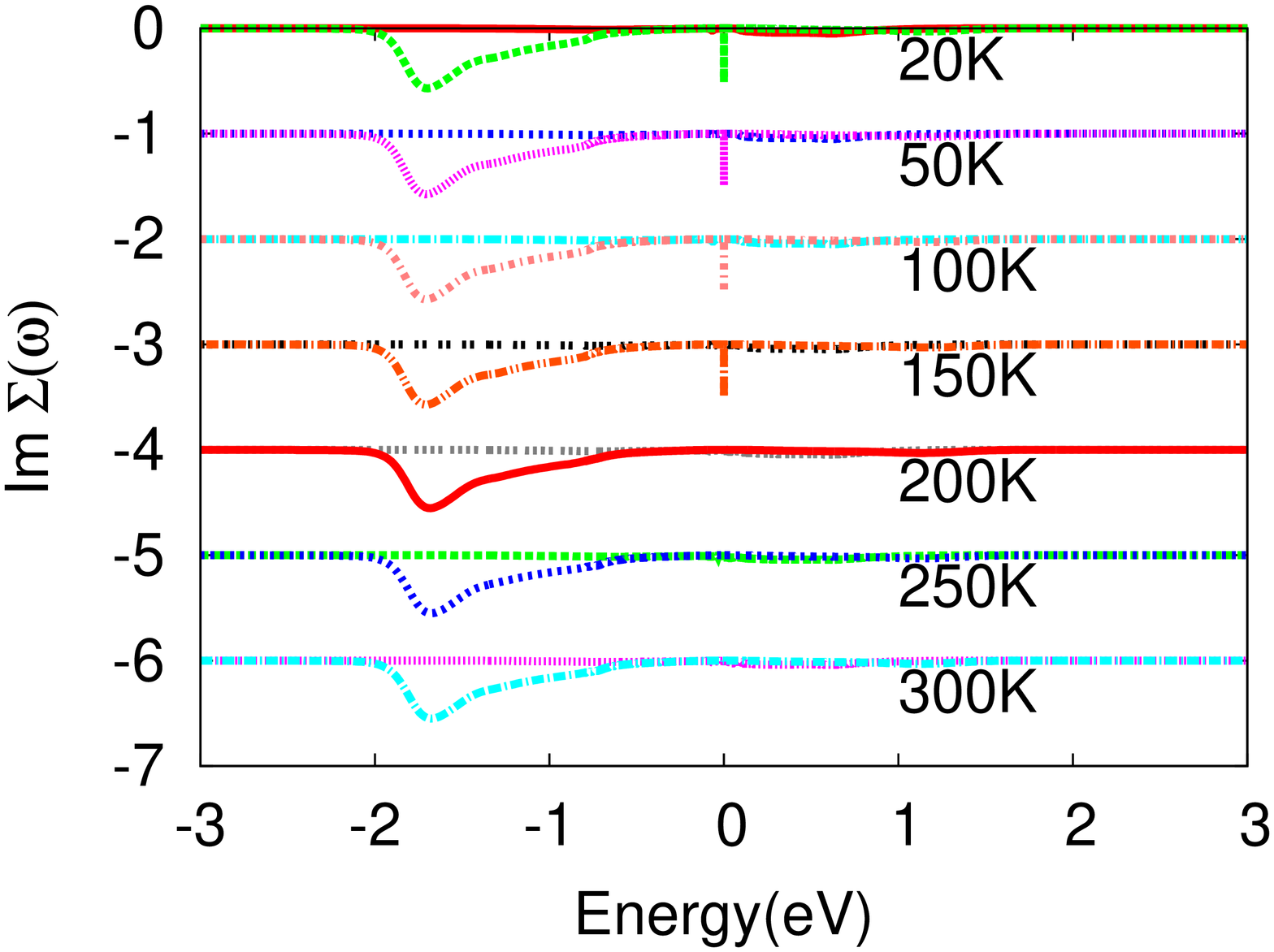}}

(c)
{\includegraphics[angle=270,width=0.6\columnwidth]{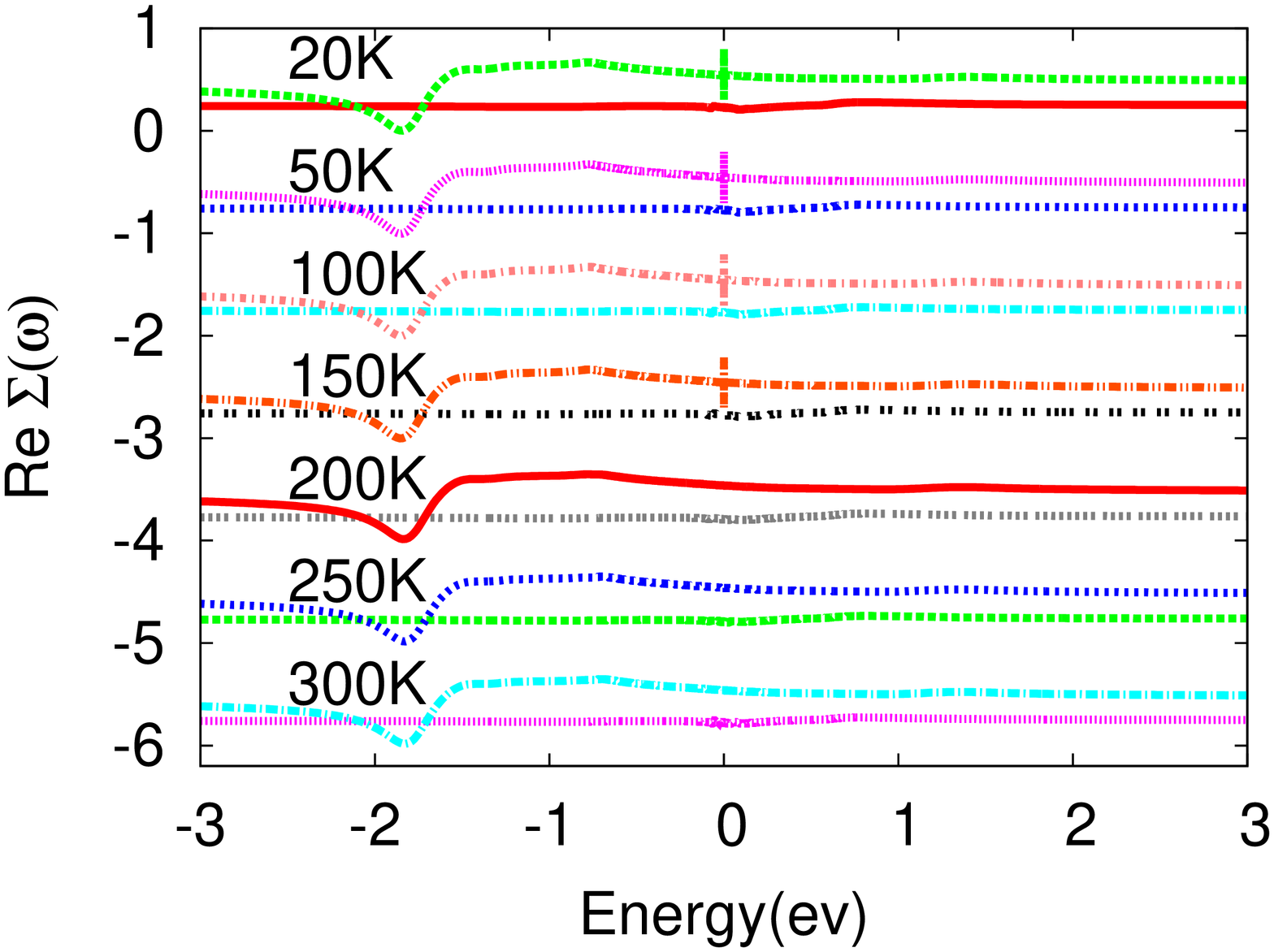}}
\caption{(Color Online) DMFT temperature evolution for (a) DOS of Ti-d and Se-p 
band (ones with peak below FL are for Se-p), (b) Imaginary 
part of self energy and (c) real part of self energy for Ti-d band and Se-p band. The self 
energies with a dip like structure close to -2 eV are for Se-p band.}
\label{fig2}
\end{figure}

We now show how the approach envisaged above gives a very good account of a 
whole range of physical responses. We show the DMFT many-body DOS for the 
`a' (Ti-d) and `b' (Se-p) bands in Fig.~\ref{fig2}a and the corresponding self
energies in Fig.~\ref{fig2}b, ~\ref{fig2}c bands at high ($T>T_{cdw}$) and
low ($T<T_{cdw}$) temperatures. 
 
The DOS shows clear `semi-metal' features,
which, however, are not those of a conventional semi-metal. Examination 
of the self-energies bares the hidden selective Mottness in the system:
Im$\Sigma_{a}(\omega,T=200$~K) clearly shows incoherent metal
features (Im$\Sigma_{a}(\omega=0)>0$ at $E_{F}(=0)$), in accord with
$(d\rho/dT)<0$ for high-$T$ (black dashed curve in Fig.2b), while the $b$-band
continues to show renormalised band insulator-like features at low energy. This
orbital-selectivity (OS) is thus related to strong normal state scattering,
rather than the onset of CDW order. Finding of OS in 1T-TiSe$_{2}$ 
is surprising (Ti being nominally $d^{0}$), but can be traced back to the fact 
that, in presence of a small number of electrons and holes induced by
$t_{ab}$, even a moderate $U_{ab} $ leads to exciton formation, and thence to
selective-Mottness in the multi-band situation that obtains in 
1T-TiSe$_{2}$. Recall that the original idea of Mott was in fact 
indelibly tied to quantum melting of correlation-induced excitons under 
pressure. Intriguingly, in 1T-TiSe$_{2}$, stabilisation of excitonic CDW order
enhances selective-Mott features at lower $T$, as emergence of
the $\omega=0$ pole in Im$\Sigma_{a,b}(\omega)$
clearly shows.  Simultaneously, however, examination of 
Re$\Sigma_{a,b}(\omega)$ also clearly shows small mass enhancements.
In extant literature~\cite{monney2}, this finding below $T_{cdw}$ is proposed as
a concrete example of a {\it reduced} effective mass due to renormalisation 
effects caused by the appearance of a strong periodic potential below 
$T_{cdw}$.  Our results are certainly in accord with this finding.  
As mentioned before, however, no large mass enhancements are theoretically 
(at least within DMFT) expected in a multiband system, even in an 
orbital-selective Mott regime.  This goes hand-in-hand with 
slight stabilisation of the normal state gap, now interpreted as a
true CDW gap. As advertised before, and in full accord with data,
sizable $T$-driven SWT accompanies the transition.
\begin{figure}
\centering
\includegraphics[angle=270,width=0.8\columnwidth]{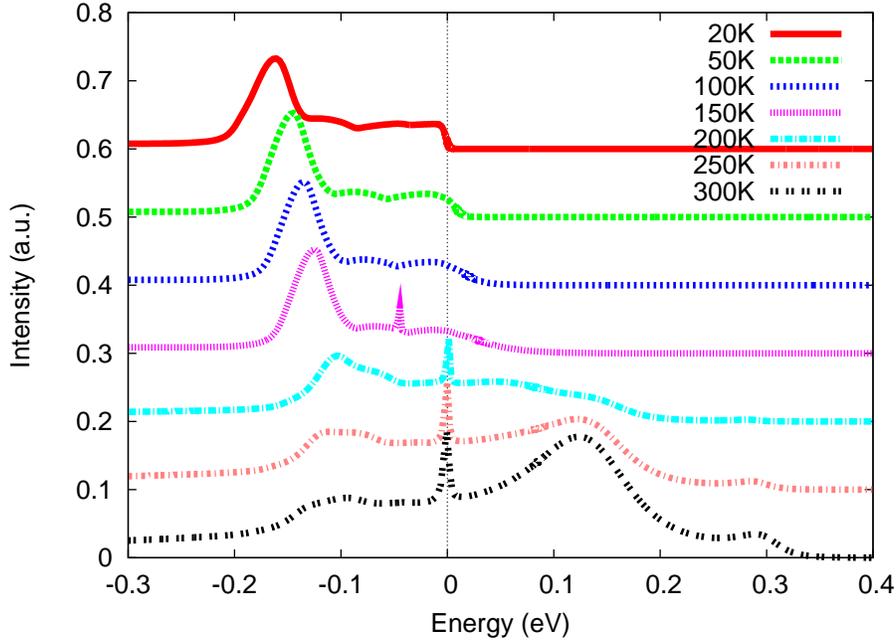}
\caption{(Color Online) Theoretical ARPES at M point at different temperatures
$(T)$.  As $T$ increases across $T_{cdw}$, the peak in the DMFT spectral 
function crosses the renormalised Fermi energy ($E_{F}$).}
\label{fig3}
\end{figure}

\begin{figure}
(a)
{\includegraphics[angle=0,width=0.45\columnwidth]{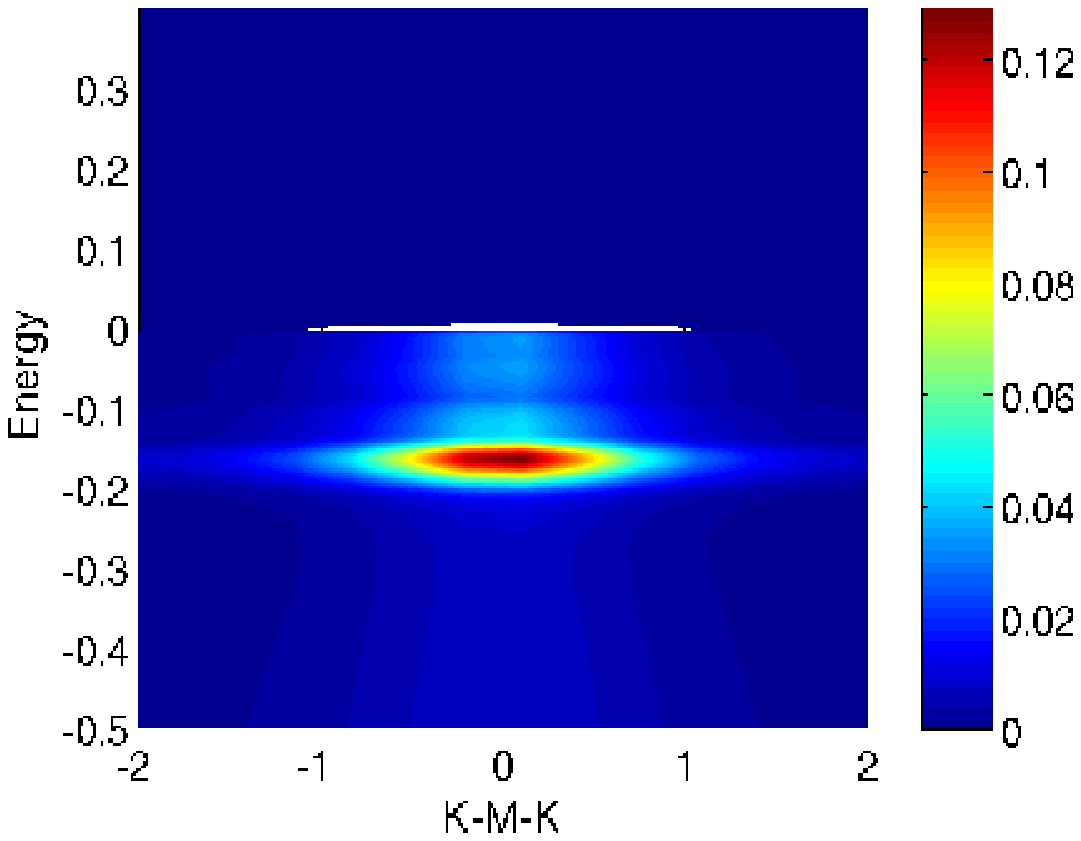}}

(b)
{\includegraphics[angle=0,width=0.45\columnwidth]{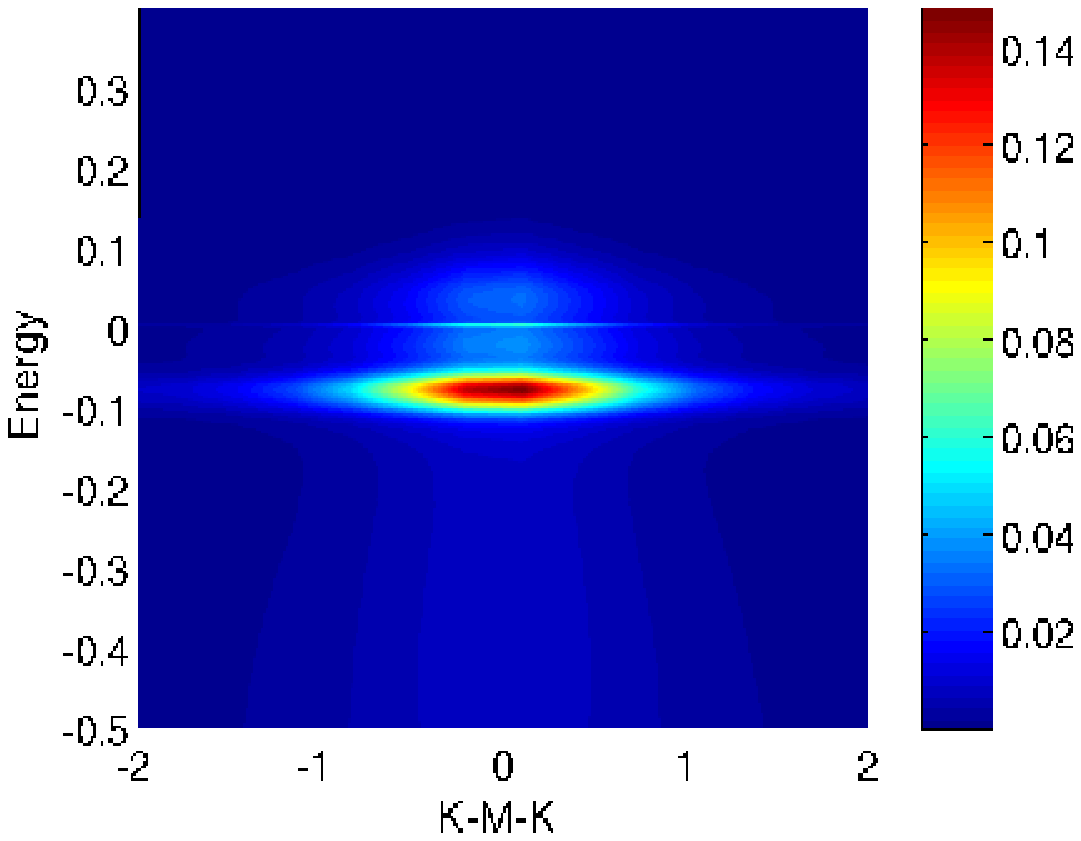}}

(c)
{\includegraphics[angle=0,width=0.45\columnwidth]{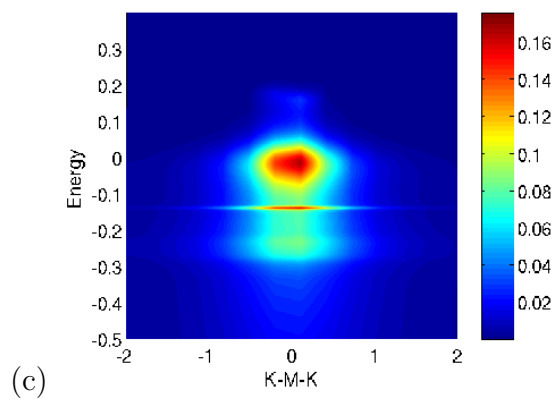}}

{\includegraphics[angle=270,width=0.45\columnwidth]{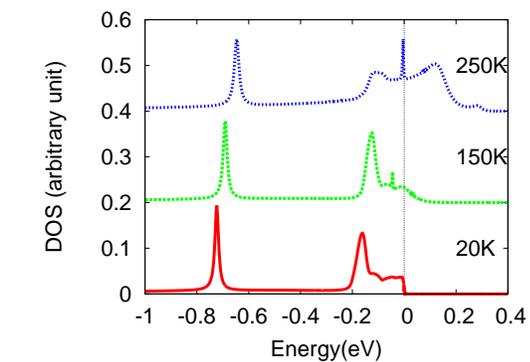}}

(d)
\caption{(Color Online) Theoretical ARPES map along K-M-K direction at 
(a) 20K (b) 150K (c) 270K (d) Total EDC at M point at different 
temperatures.  Excellent accord with ARPES data is clearly seen.}
\label{fig4}
\end{figure}

If the PEL alternative is to be credible, the full range of
observations must be explicable without any additional
assumptions. In Fig.~\ref{fig3} we compare our DMFT
one-particle spectral function, $A_{a,b}({\bf
k},\omega)=-$Im$G_{a,b}({\bf k},\omega)/\pi$, and renormalised
band dispersion, $E_{{\bf k},a,b}=\epsilon_{{\bf k},
a,b}+$Re$\Sigma_{{\bf k},a,b}$, with ARPES data~\cite{aebi}
where these exist.  This is already a stringent test for theory: while
LDA plus static-HF can conceivably yield
agreement with band {\it dispersions}, the real test is a
simultaneous description of the ARPES lineshapes.  For a dynamically
fluctuating `liquid', the latter is expected to show broad
continuum-like features without
infra-red (LFL quasiparticle) poles. Rather remarkably, our LCAO+DMFT
results provide an excellent
semi-quantitative description of extant ARPES dispersions and
lineshapes up to rather high energy.  In particular,
they bare the preformed excitonic features in $E_{{\bf k},a,b}$
and clear, associated `gap' features in ARPES lineshapes above
$T_{cdw}$. Excellent accord in all details, including the band
positions, their intensity
distributions, and band-shifts as a function of $T$, with the ARPES
dispersion is clear from a direct
comparison between our Fig.~\ref{fig4}(a-c) with Fig.2 of Monney {\it
et al.}~\cite{aebi}  In particular, we can even identify the
detailed features in the $T$-dependence of the ARPES intensity 
(Fig.~\ref{fig3}) with data: $(i)$ the peak at $\simeq -0.2$~eV is identified
as the valence band backfolded to the $M$ point by comparing the red curve
in Fig.~\ref{fig3} with the top panel in Fig.~\ref{fig4}. $(ii)$ a new peak,
labelled `$C$' by Monney {\it et al.}, also appears below $T_{cdw}$ and gets
more pronounced upon decreasing $T$, precisely as seen. $(iii)$ The peak 
`$D$', identified as a `quasiparticle peak' originating from coupling to
phonons in experiment is also obtained in DMFT in very good qualitative 
accord with data: especially interesting is that it is quickly damped out as
$T$ increases, exactly as seen in the top four curves in our Fig.~\ref{fig3}.
However, at high $T$, we resolve an additional (sharper) low-energy peak, also
arising from electron-phonon coupling, which remains dispersionless above 
$T_{cdw}$: this feature is not seen in the ARPES results of Monney 
{\it et al.}  Finally, we also cannot observe peak `$A$', identified with a
second, spin-orbit split, valence band, since we have not used the full set
of Se-$p$ and Ti-$d$ bands in this work.  Nevertheless, the accord between 
theory and experiment is excellent with regard to features relevant for the 
CDW in 1T-TiSe$_{2}$.

We now consider the ARPES `band structure' maps. At high-$T$, along K-M-K direction in the 
Brillouin zone the putative valence band, (VB) heavily damped by strong (due to preformed 
excitons) scattering, crosses $E_{F}(=0)$ while the putative conduction band 
(CB) lies totally above $E_{F}$.  This conflicts with the LDA (or LCAO, see Fig.~\ref{fig1}) 
results, which predict a sizable overlap between VB and CB states at $E_{F}$, and 
reflects the failure of the FS nesting mechanism in the {\it real} one-particle spectra (in 
fact, to cure such conflicts, the standard explanation within conventional views has been 
to attribute such discrepancies between LDA and ARPES data to uncontrolled excess~
\cite{monney2} of Ti. Our view naturally produces this, without having to take recourse 
to such extraneous conditions). Clear Hubbard-band shake-up features  in DMFT are 
also seen in the color-plot: in particular, we predict that the $\omega >0$ part of the 
lineshape (at M point in Fig.~\ref{fig4}d) should lend itself to observation in inverse 
ARPES (ARIPES) studies in future. In addition, comparison of the DMFT energy distribution 
curve (EDC) with extant ARPES result at the M point also shows (Fig.3) very good accord as 
a function of $T$ over the whole range. A moderate band narrowing of LCAO bands and 
broad spectral lineshapes with appreciable $T$-dependent SWT in one single system are 
generic fingerprints of strong dynamical correlations in multi-orbital systems, and so, as 
alluded to before, the second feature above, essential for describing `liquid' correlations, 
cannot be accounted for by a static-HF theory, as done so far~\cite{wezel}. 

Detailed Fermi surface (FS) maps as a function of $T$ in 
1T-TiSe$_{2}$ are rare. We have taken the FS 
mapped out by Rossnagel {\it et al.}~\cite{rossnagel} to study how the 
PEL idea survives this important test. 
While instabilities driven by LDA FS forms the backbone of weak 
coupling or itinerant views, it is also possible, 
in a renormalised itinerant or Mottness based theories, 
that these could involve a {\it new} FS sizably 
reconstructed by correlations.  That the FS does not reconstruct 
across $T_{cdw}$ in 1T-TiSe$_{2}$ was one of the main 
(among others, see above) arguments for invoking unconventional PEL 
scenarios in the first place.  Close inspection 
of FS evolution across $T_{cdw}$ shows that, while this is undoubtedly 
correct, there are still specific features 
which any theory needs to confront: (i) the band pockets are 
smeared out at high temperature ($T>T_{cdw}$) and, more 
importantly, a much brighter ring structure appears at the M point 
below $T_{cdw}$.  

\begin{figure}

(a)
{\includegraphics[angle=360,width=0.5\columnwidth]{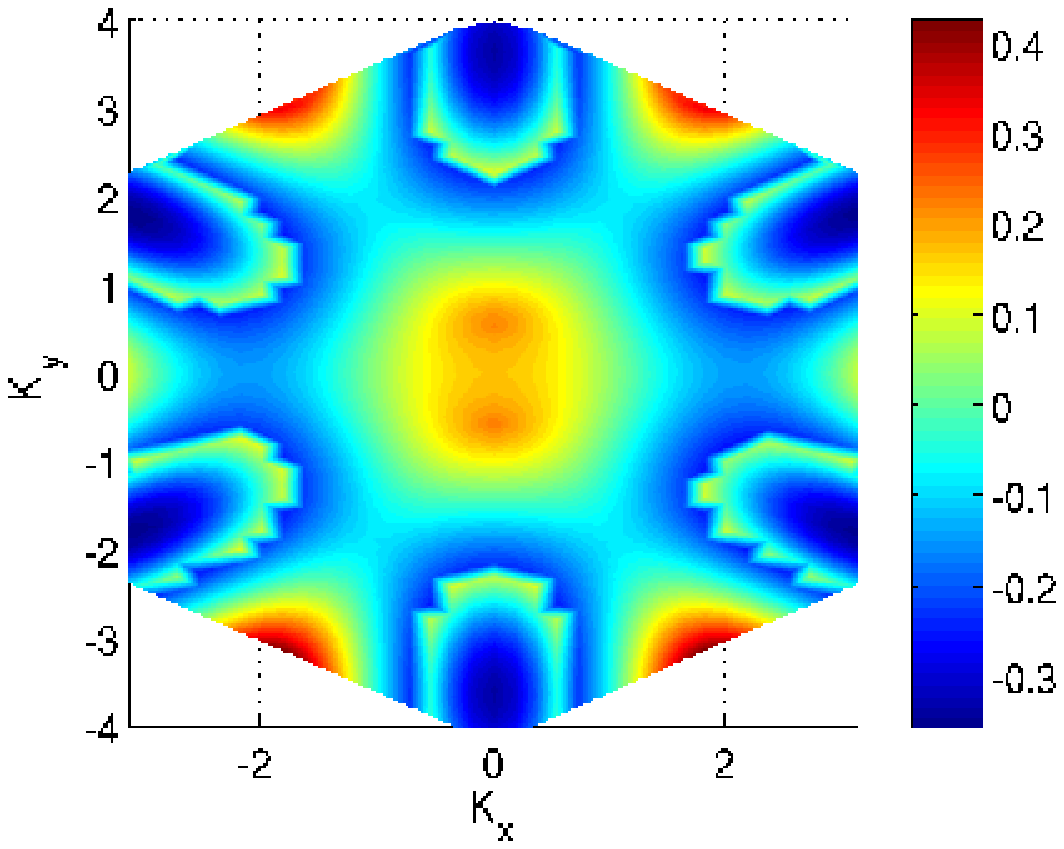}}

(b)
{\includegraphics[angle=0,width=0.5\columnwidth]{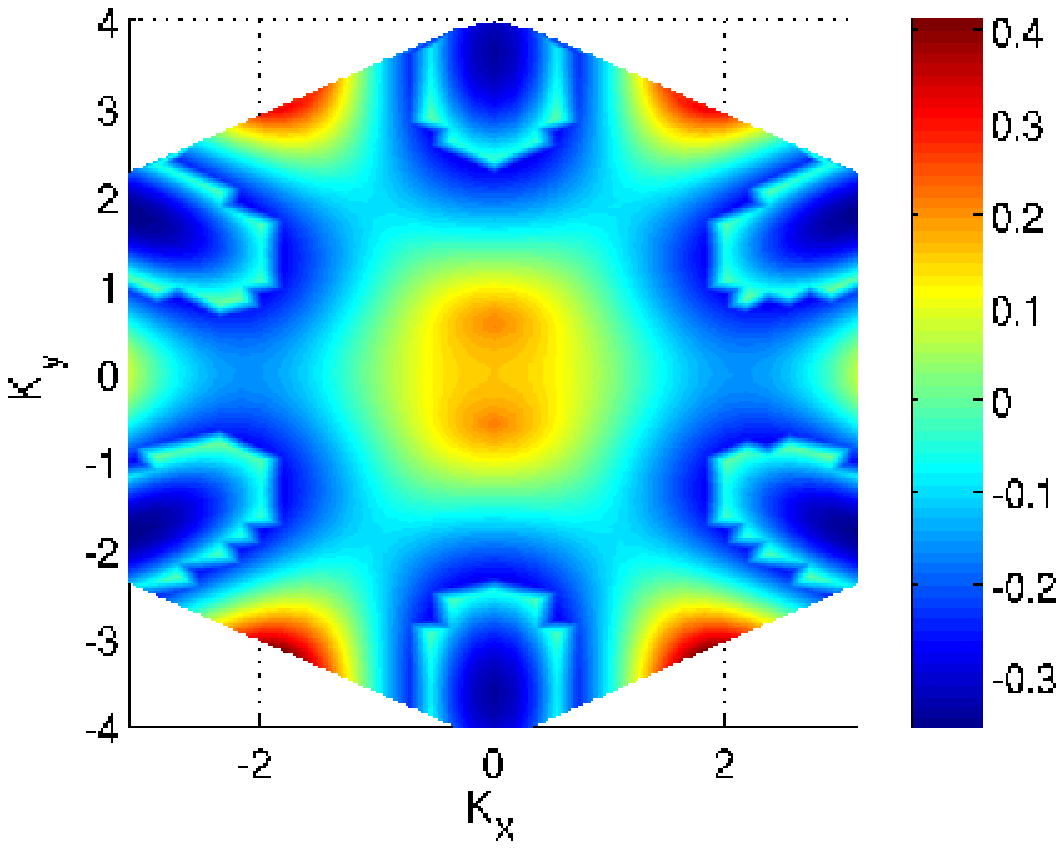}}

(c)
{\includegraphics[angle=0,width=0.5\columnwidth]{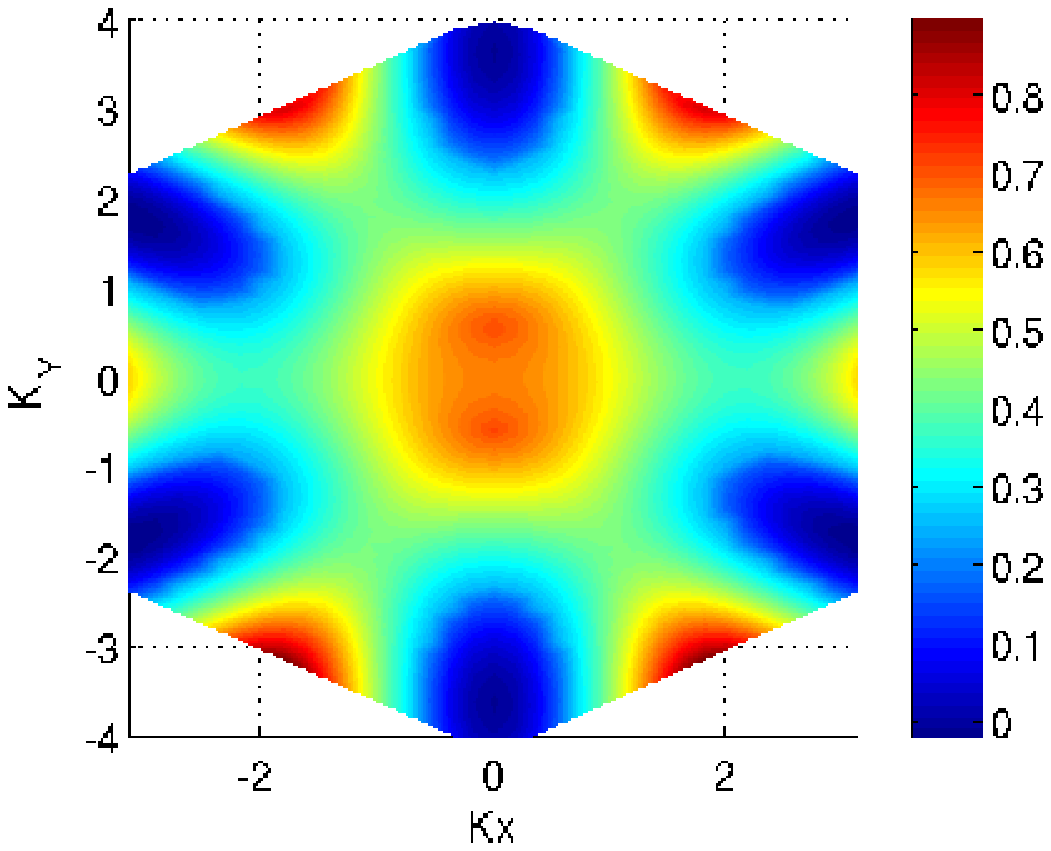}}
\caption{(Color Online) DMFT FS map at (i) 20K (ii) 150K (iii)
270K, showing very good agreement with the T-evolution of the Fermi surface 
in ARPES data of Rossnagel {\it et al.}~\cite{rossnagel}}
\label{fig5}
\end{figure}

In Fig.~\ref{fig5}, we show 
our DMFT FS at low (top panel) and high-$T$ (bottom panel). As expected due
to strong PEL scattering, the high-$T$ FS is
sizably smeared out (arises from the finite
Im$\Sigma_{a}(\omega=0)$ at 200K from Fig.~\ref{fig2}). Remarkably
enough, we {\it also} find a clear ring feature at the M point below
150K (lower panels), in remarkable accord with
data.  Since this ring-like feature becomes well-defined only below
$T_{cdw}$, it is a direct consequence of CDW
order-induced reconstruction of electronic states. This is fully
consistent with the excitonic CDW view, where
CDW follows excitonic `solid' order, and the latter arises principally via
interplay between $t_{ab}^{\bf k}$ and $U_{ab}$, implying backfolding of Se
$p$ band from the $\Gamma$ point due to the 2x2x2 CDW superstructure
formation. Further, close observation of DMFT results shows that the central 
pocket around the $\Gamma$ point has seemingly developed `elongated' shape
instead of the hexagonal shape expected from the LCAO result.  We believe that
this important modification, hitherto not noted sufficiently, arises due to 
an orbital-dependent electronic structure reconstruction that is 
essentially driven by the ${\bf k}$-dependent form factor of the inter-orbital
hybridisation ($t_{ab}({\bf k})$). Experimental confirmation of this feature
would thus constitute additional support for a PEL scenario. Finally, the
stabilisation and slight increase of the normal state `gap' below $T_{cdw}$ also 
accords with optical data~\cite{opt} and with the semi-metal-to-semiconductor
characterisation of the normal-CDW transition in 1T-TiSe$_{2}$.

Thus, such quantitative agreement between our excitonic-DMFT results 
premised upon a novel PEL view and
ARPES in all important details lends strong credence to the idea of a
dynamically fluctuating excitonic liquid at high $T$ giving way to a 
low-$T$ CDW order. However, to further qualify as a credible 
candidate, the {\it same} formulation must also describe 
transport as well. Fortunately, in DMFT, this task is simplified: it is 
an excellent approximation to compute transport co-efficients directly from 
the DMFT propagators $G_{a,b}(k,\omega)$~\cite{silke}, since
(irreducible) vertex corrections rigorously vanish for one-band models, and
turn out to be surprisingly small even for the multi-band case. 

\begin{figure}
\centering
\includegraphics[angle=270,width=0.8\columnwidth]{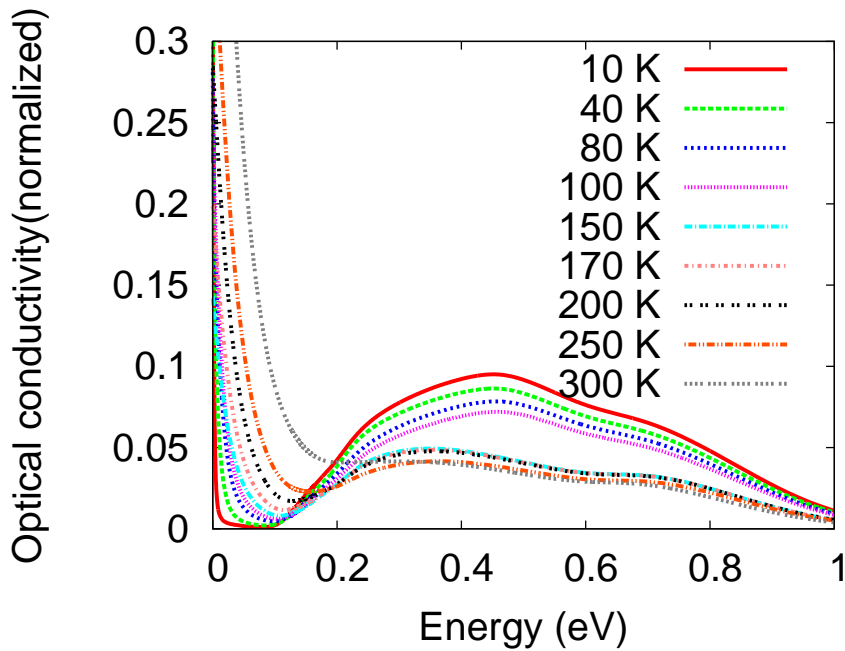}
\caption{(Color Online) DMFT results for $T$-dependent optical conductivity,
showing good agreement with optical data across $T_{cdw}$.}
\label{fig6}
\end{figure}
In Fig.~\ref{fig6}, we show the optical conductivity $\sigma(\omega)$ as 
a function of $T$, wherein very good accord with data up to an energy 
$\simeq O(0.8)$~eV is clear. A clean CDW gap at low $T$ closes rapidly with 
increasing $T$ via rapid spectral weight transfer from the relatively sharply defined hump 
at $0.4$~eV to the infra-red regions, precisely as seen. Given that we have kept 
only the two bands crossing $E_{F}$ in LCAO, agreement at higher energy 
($\ge 0.8$~eV) is not expected. The relative sharpness of 
the low-energy optical response at low $T$ is deceptive: it is not a FL Drude 
peak, and in fact, (fully consistent with the selective-Mott behavior in
the DMFT spectra above) no FL coherence sets in, even at lowest ($10$~K)
$T$ in DMFT.  It is rather a reflection of reduced incoherence due to
CDW gap opening.  Finally, we also resolve an isosbectic point in
$\sigma(\omega,T)$ curves (where $\sigma(0.2 eV)$ remains invariant) 
at different $T$: this is clear manifestation of sizable dynamical 
correlations~\cite{arghya}, and is again in good accord with the isosbectic 
point seen around $0.27$~eV in the optical study~\cite{opt}. 

\begin{figure}
\centering
\includegraphics[angle=270,width=0.8\columnwidth]{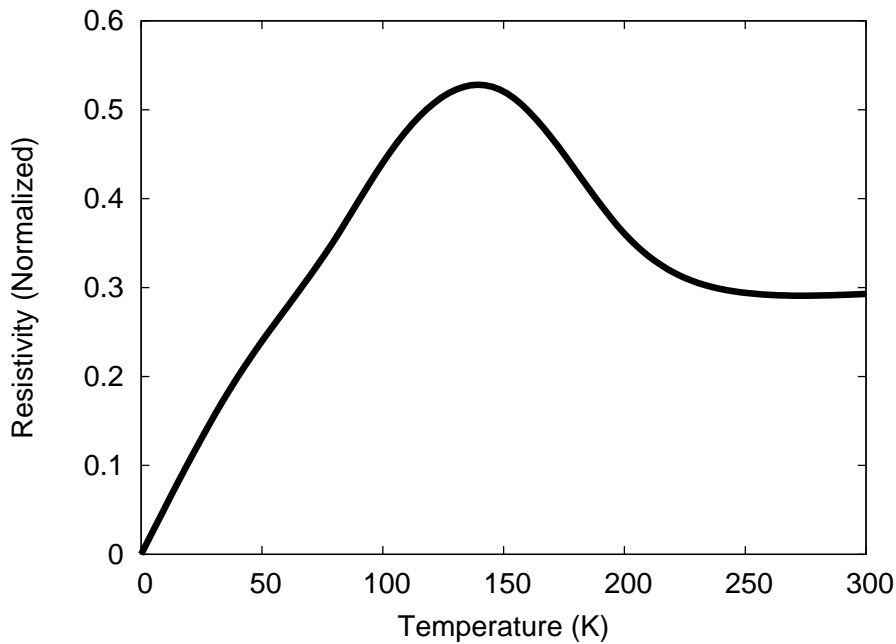}
\caption{$T$-dependent dc resistivity within DMFT.  Except for the low-$T$ part, LCAO+DMFT results agree very well with experimental results, 
including a broad maximum, rather than any sharp non-analytic change, across
$T_{cdw}$ (see text). }
\label{fig7}
\end{figure}

The DMFT
resistivity (Fig.~\ref{fig7}) also shows an insulator-like behavior
above $T_{cdw}$, a broad peak without
any anomaly below $T_{cdw}=200$~K, and bad metallic behavior below $T_{cdw}$, 
in full accord with data~\cite{sipos}.  The insulator-like
behavior for $T>T_{cdw}$ is a reflecting of the large normal state incoherence,
whence carriers strongly scatter off fluctuating, incoherent
(preformed) excitons.  The bad-metallic behavior below $T_{cdw}$ is then 
naturally attributable to reduction of this strong normal state scatttering
due to $(i)$ opening up of a CDW gap, and $(ii)$ concomitant increase in 
the tendency to excitonic coherence.  Given the bad-metallic resistivity, the 
consequent 
short scattering mean-free path (in fact, $k_{F}l\simeq O(1)$) invalidates a 
quasiclassical Boltzmann equation-based approach to transport.  Interestingly, 
this is precisely the regime where DMFT should work best.  The absence of 
a sharp ordering anomaly at $T_{cdw}$ is additional 
evidence against a weak coupling view of the instability. In fact, in a 
weak-coupling instability, resistivity should have shown a sharp ordering anomaly 
at $T_{cdw}$ on general grounds~\cite{langer} (in addition, $d\rho/dT$ must also show 
critical behavior, with exponents linked to those extracted from thermodynamic
measurements). The strong coupling view is further supported by finding of a 
large $2\Delta/k_{B}T_{cdw}\simeq 7-10$ in TMD (about $7$ for 1T-TiSe$_{2}$ 
and 10 for 2H-TaSe$_{2}$~\cite{mcmillan}). This feature is reminiscent of 
high-T$_{c}$ cuprates~\cite{juan} and implies that CDW formation is associated
with a BEC, rather than a BCS-like scenario for the exciton instability.
Also, strong inelastic scattering needed to rationalise bad-metallicity 
above $T_{cdw}$ and sizable $T$-induced SWT are characteristic signatures of a 
strong coupling limit. 

Thus, taken together, very good accord with ARPES 
and transport data strongly supports our basic hypothesis: $(i)$ the normal 
state is a strongly fluctuating liquid of incoherent excitons, and $(ii)$ CDW 
order in 1T-TiSe$_{2}$ must fall into the strong coupling class, 
qualitatively different from a conventional Overhauser transition of 
well-defined band (LFL) quasiparticles.

Emergence of CDW order from a PEL should leave further specific signatures in 
other data. First, CDW and excitonic correlations must now track each other 
beyond $T_{cdw}$, well into the PEL state. 
\begin{figure}
(a)
{\includegraphics[angle=270,width=0.7\columnwidth]{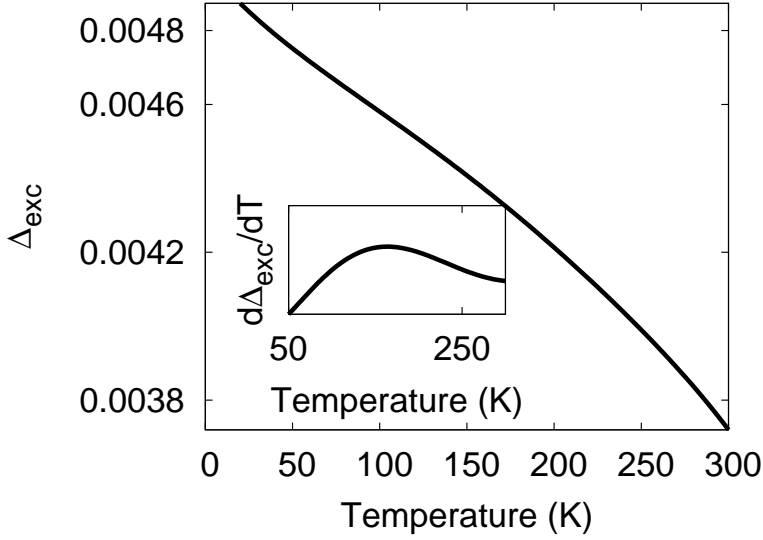}}

(b)
{\includegraphics[angle=270,width=0.7\columnwidth]{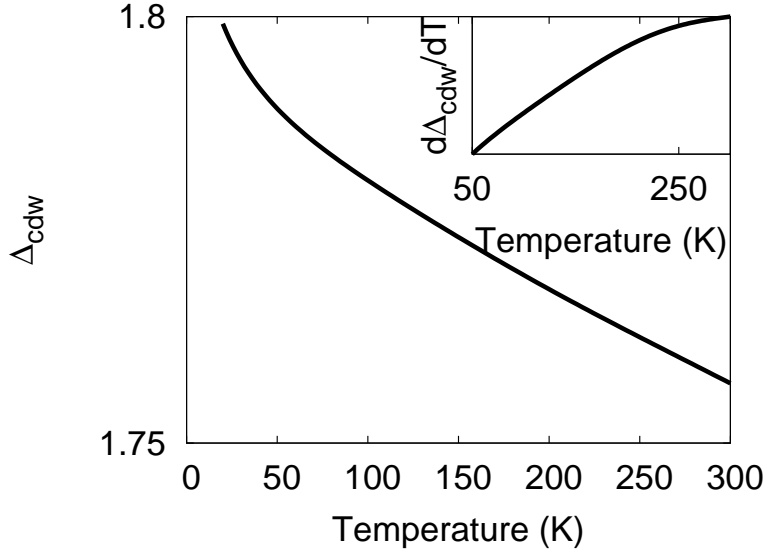}}
\caption{(a)Expectation values and derivative (inset) of 
inter-orbital excitonic (Im$G_{ab}$), and of 
(b) CDW order parameter, n$_a$-n$_b$ = $\Delta_{cdw}$ and its derivative 
(inset) as a function of $T$.}
\label{fig8}
\end{figure}
In Fig.~\ref{fig8}, we show the
excitonic and CDW order parameters, $\Delta_{exc}\propto \langle \Gamma^{x}\rangle=\langle
(c_{ia}^{\dag}c_{ib}+h.c)\rangle, \Delta_{cdw}\propto \langle \Gamma^{z}\rangle=
\langle n_{a}-n_{b}\rangle/2$, along with their $T$-derivatives, as a function 
of temperature ($T$). In particular, as expected in the PEL scenario, {\it both} follow each
other and are finite way above $T_{cdw}$, in nice qualitative accord with the 
$T$-dependence of the CDW order parameter from ARPES~\cite{aebi}. The absence of a 
conventional (BCS-like) mean-field transition can be rationalised by noticing that, in 
presence of the el-lattice coupling, $H_{el-l}=g\sum_{i}(A_{i}+A_{i}^{\dagger})
(c_{ia}^{\dag}c_{ib}+h.c)$, the universality class of the normal-CDW 
transition turns out to be that of an Ising model in 
an external field. This implies that the (mean-field) transition is smeared 
into a smooth crossover.  The relevance of excitonic correlations is also 
visible as follows.  We see that $\rho(T)$ closely follows the $T$-dependence
of $d\Delta_{exc}/dT$ above (Fig.~\ref{fig8}), but does not show any obvious 
correlation with $d\Delta_{cdw}/dT$: this shows that the anomalous resistivity is 
caused by carriers scattering off incoherent excitonic (liquid-like) correlations. In
our PEL scenario, the change to metallic behavior in $\rho(T)$ is attributable 
to reduction of the strong scattering when excitons condense at $T_{cdw}$. 
True CDW order now follows a true BEC of preformed excitons at lower $T$.  

\begin{figure}
\centering
{\includegraphics[angle=270,width=0.7\columnwidth]{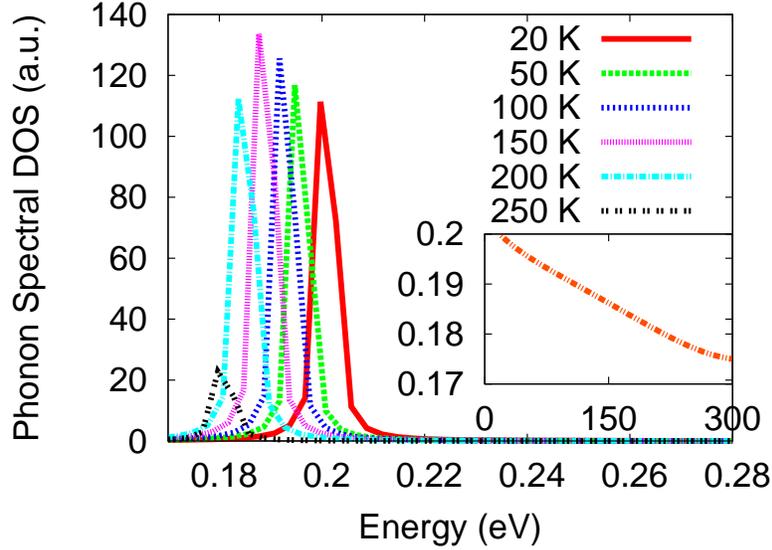}}
\caption{(Color Online) Phonon spectral function at different temperatures. 
The inset shows change in peak position of the $A_{1g}$ phonon mode with $T$,
in good accord with Raman data above $T_{cdw}$ (see text).}
\label{fig9}
\end{figure}

Given the specific form of the electron-phonon coupling, now also interpretable 
as coupling of interband excitons to $A_{1g}$ phonon mode by symmetry, the 
lattice is expected, very generally, to react to the $T$-dependent 
changes (incoherent at high $T$, more coherent at lower $T$) in 
exciton dynamics. In particular, the DMFT phonon spectral function, 
computed from $\rho_{ph}(\omega)=(-1/\pi)$Im$D(\omega)=(-1/\pi)$Im$
[2\Omega_{0}/(\omega^{2}-\Omega_{0}^{2}-2g^{2}\Omega_{0}\chi_{ab}(\omega))]$
and shown in inset of Fig.~\ref{fig9}, should mirror excitonic CDW
correlations. 

Several interesting features stand out: (i)
$\rho_{ph}(\omega)$ shows maximum intensity with reduced linewidth 
around $T_{m}=150$~K, precisely where the resistivity peaks.  (ii)
Above $T_{cdw}$, the phonon spectrum is noticeably broader, reflecting
coupling to incoherent excitonic liquid modes, and (iii) asymetry in
$\rho_{ph}(\omega)$ {\it increases} as $T$ is lowered
below $T_{cdw}$.  This reflects precursors of a Fano-like structure,
arising from treatment of $H_{el-l}$ beyond
the adiabatic limit within our DMFT.  Onset of excitonic coherence
below $T_{cdw}$ reduces strong normal state scattering, sharpening the 
phonon spectrum.  (iv) Finally, we find that the detailed $T$-dependence 
of both the $A_{1g}$-phonon lineshape and linewidth excellently matches 
that extracted from a Raman scattering study~\cite{kidd}. 
Only below $T_{cdw}$ is there a discord between measured and computed 
linewidths: it levels off to a constant in experiment, a feature not captured 
by our present calculation. The reason is that we have not carried out a study 
of the change in phonon dynamics below $T_{cdw}$. A full study of lattice 
effects must rest on a more realistic input to the full phonon spectrum of 
1T-TiSe$_{2}$, a point we leave for future study.
Thus, taken together, very good quantitative accord with a whole host
of spectral and transport responses for 1T-TiSe$_{2}$ {\it and} a 
comprehensive qualitative rationalisation of structural features in 
one single theoretical picture constitutes overwhelming support for a novel 
PEL view. Thus, the central conclusion of our work is:  
{\it The CDW instability must now be interpreted as a strong
coupling instability of an incoherent PEL normal state, rather 
than as a weak coupling Fermi surface nesting instability of a 
good LFL}.

The above findings have important implications for SC arising under pressure. 
A finite pressure increases $a-b$ band overlap via $t_{ab}$,
leading to a redistribution of electrons and holes, and weakens
excitonic CDW order. From Fig.~\ref{fig8}a, this implies enhancement of 
excitonic fluctuations, also seen by noticing that a fall of
$\langle \Gamma_{i}^{z}\Gamma_{j}^{z}\rangle$ must transfer weight to the transverse 
part, i.e, enhance $\langle (\Gamma_{i}^{+}\Gamma_{j}^{-}+h.c)\rangle$ via the 
pseudospin sum-rule ${\bf \Gamma}^{2}=\Gamma(\Gamma+1)$. Thus, near the critical $p=p_{c}$ 
where CDW is destroyed, we expect a maximisation of excitonic fluctuations. 
These can constitute the critical collective (electronic) fluctuations
that could lead to an instability to an inter-band SC (competing with CDW) 
with a finite $\Delta_{ab}\simeq \langle c_{ia\sigma}c_{jb-\sigma}\rangle$.
However, normal state incoherence implies that SC must now arise
directly from the incoherent metal via a two-particle instability: this can 
be explored by solving $H=H_{el}+H_{res}$ in the pair channel. The el-lattice 
coupling will further enhance the effective pair interaction
in $H_{res}$ above~\cite{arghya}. Even without the benefit of a detailed DMFT
calculation for the SC phase, we thus expect the SC $T_{c}$ to be maximal 
close to $p_{c}$ if dynamical excitonic fluctuations mediate cooper pairing: 
this is again in accord with observations~\cite{sipos} and the causal 
link is compelling.

We believe the present work also has further important implcations on a 
broader level.
Our unexpected finding of OS bad-metallicity in the high-$T$ PEL brings 
1T-TiSe$_{2}$ closer to the bad actors in $d$- and $f$-band systems, where 
OS is a consequence of orbital-selective Hubbard correlations~\cite{f-qcp} 
(or, in the cuprates, momentum-selective correlations~\cite{juan}). Given that 
1T-TaS$_{2}$ has long been known to be a Mott insulator, our proposal 
of selective-Mottness and associated excitonic liquid features 
is quite reasonable, albeit novel.  However, since magnetism is never 
an issue in the TMDs of interest, what underpins anomalous responses reminiscen 
 of $d$- and $f$-band critical systems near magnetism is moot. The common 
unifying mechanism seems to be selective-Mottness, which is 
indeed one of the main contenders for understanding anomalous quantum 
criticality in $d$ and $f$-band systems~\cite{senthil,sachdev}.  Selective-Mottness itself can always occur via different mechanisms: in particular, depending upon the microscopics of a given system, it can be associated with magnetism, 
but it clearly does not always need to be so.  It can also be associated with 
other density-wave instabilities in the charge or orbital sectors in mutliband
systems such as TMDs.  Nevertheless, a common element, namely destruction of 
Landau FL quasiparticles, would always accompany onset of selective-Mottness.  
Thus, in light of the present work, it should not be too surprising that the 
high-$T$ PEL bears some similarities with the FL$^{*}$ theory~\cite{senthil} in 
the $f$-electron QCP context. Evidence attesting to this comes from the 
fact that only the Se-$p$ band crosses $E_{F}$, while the Ti-$d$ band lies 
above $E_{F}$ already in the high $T$ liquid. In other words, the observed FS 
has already reconstructed from its LDA counterpart to reflect the preformed 
excitonic character of the normal state. Spectral and transport features show 
characteristic incoherence features, as expected in an FL$^{*}$ state, and the 
efficacy of DMFT as a valuable tool to understand these is also known in the 
$f$-electron context~\cite{kotliar,f-qcp}. That this identification implies 
an interesting scenario, similar to those invoked for anomalous metals on 
the border of ($T=0$) magnetism may also permit a rationalisation of the
apparent `similarities' claimed to exist between under-doped cuprates and 
TMD~\cite{bsenko}. Viewed from the perspective of our study,
these similarities are, ultimately, manifestations of the
selective-(bad) metallicity in both cases: momentum-selective
in under-doped cuprates, and orbital-selective in 1T-TiSe$_{2}$. Thus, we
arrive at a perhaps general idea with broader
appeal across classes of correlated systems:  Selective Mottness
(whether orbital or momentum) induces
critical electronic liquid states characterised by loss of LFL
coherence. This collectively fluctuating
electronic fluid can subsequently become unstable, either to a myriad
(charge, spin, orbital) of density wave
orders, or to (competing) unconventional superconductive orders,
depending upon the actual microscopics of the system under study.  
In TMD, material specific reasons favor competing inter-band
CDW and SC orders from such a strongly fluctuating excitonic liquid.

\section{Conclusion}
In conclusion, we show that a whole host of physical responses in
1T-TiSe$_{2}$, difficult to reconcile with a band FS nesting or 
band-JT mechanisms, is provided a natural and good quantitative 
explication within a new PEL alternative. Finding of novel, 
orbital-selective Mott features in renormalised normal state 
electronic structure via DMFT leads to an excellent description of a range 
of spectral and transport data in both normal and CDW states. More 
importantly, it brings the TMD closer to the anomalous critical metals 
of much more recent interest.  Along with its success for 
2H-TaSe$_{2}$~\cite{arghya}, our 
work relates the PEL idea to a more generic theoretical level for TMD. 
Particularly interesting should be to test how this new proposal fares for 
the even more correlated 1T-TaS$_{2}$~\cite{sipos1}, which is undoubtedly 
a known Mott insulator~\cite{tosatti}, as well as competition between 
CDW and SC, in future work.\\

SK acknowledges CSIR (India) for a senior research fellowship.  MSL thanks the
ILL Grenoble for financial support and hospitality. We thank H Cercellier for 
very helpful discussions.

\section{Appendix}

\subsection{Tight binding band structure}

TiSe$_2$ is a layered material with hexagonal layers of Ti sandwiched 
between layers of Se atoms: a 1T-polytype transition metal dichalcogenide. 
The Ti d-band is in a d$^0$ state in TiSe$_2$. Due to octahedral 
coordination the Ti atom d-shell is split into a set of high energy e$_g$ 
orbitals and low lying, degenerate t$_{2g}$ orbitals. 
We consider only the low lying t$_{2g}$ orbitals, for, they are the ones
that form the hybridized bands with Se-p orbitals close to the Fermi level. 
In the LCAO calculation, therefore, we consider charge transfer between 
them and the surrounding Se 4p orbitals, so that six Se p orbitals and three 
Ti d orbitals per unit cell are involved.  Our results are similar to those of 
Wezel {\it et al.}~\cite{wezel}, but we do not invoke quasi-1D features 
resulting from the orbital dependence of the hopping matrix elements as the
driving cause of CDW.  However, the shape of the central pocket around the 
$\Gamma$ point in the DMFT Fermi surface in the main text does hint 
toward such anisotropic features getting more pronounced in the CDW 
ordered phase.

The tight binding Hamiltonian is thus constructed as

$$\hat{H}=\sum_{i,\alpha}\frac{\Delta}{2}(\hat{d}^{\dagger}_{i,\alpha}
\hat{d}^{\phantom \dagger}_{i,\alpha} - \hat{p}^{\dagger}_{1i,\alpha}
\hat{p}^{\phantom\dagger}_{1i,\alpha}-\hat{p}^{\dagger}_{2i,\alpha}
\hat{p}^{\phantom \dagger}_{2i,\alpha}) $$\\ 
$$+ \sum_{<i,j>,\alpha,\beta}(t^{dd}_{\alpha,\beta,i-j}
\hat{d}^{\dagger}_{i,\alpha}\hat{d}^{\phantom \dagger}_{j,\beta} + 
t^{pp}_{\alpha,\beta,i-j}[\hat{p}^{\dagger}_{1i,\alpha}\hat{p}^{\phantom 
\dagger}_{1j,\beta}-\hat{p}^{\dagger}_{2i,\alpha}\hat{p}
^{\phantom \dagger}_{2j,\beta}]) $$\\ 
$$+\sum_{<i,j>,\alpha,\beta}(t^{pd}_{\alpha,\beta,i-j}
[\hat{d}^{\dagger}_{i,\alpha}\hat{p}^{\phantom \dagger}_{1j,\beta} + 
\hat{d}^{\dagger}_{i,\alpha}\hat{p}^{\phantom \dagger}_{2j,\beta} + 
H.c.]+$$\\
$$t^{pp}_{\alpha,\beta,i-j}[\hat{p}^{\dagger}
_{1i,\alpha}\hat{p}^{\phantom \dagger}_{2j,\beta}+H.c.]). \eqno(5.1.1)$$ 

Here $\hat{d}^{\dagger}_{i}, \hat{p}^{\dagger}_{1i}$ and 
$\hat{p}^{\dagger}_{2i}$ create electrons on the Ti, the upper Se and 
the lower Se atom respectively. The labels $\alpha$ and $\beta$ run over 
all possible orientations of the Ti d-t$_{2g}$ and Se p orbitals and 
$<i,j>$ denotes neighbouring sites. $t^{dd}, t^{pp}$ and $t^{pd}$ are 
the different hopping matrices for d and p orbitals and $\Delta$ is the 
chemical potential.

This Hamiltonian results in a 9$\times$9 matrix using Slater-Koster 
integrals, and several matrix elements vanish rigorously due to crystal 
symmetry. We obtain 
the LCAO band structure by diagonalizing the resultant matrix. By adjusting 
the values of Slater-Koster integrals and the chemical potential the tight 
binding bands can be fit to the extant LDA calculation. Setting $dd\sigma=
-0.2, dd\pi=0.2, dd\delta=-0.5, pd\pi=0.5, pp\sigma=0.4$ and $\Delta=2.0$ we 
get a TB fit in good quantitative agreement with 
the LDA results~\cite{yoshida,jishi}. 

\subsection{Electron phonon Self Energy}

To calculate electron-phonon self-energy we incorporate the procedure 
first used by Ciuchi et al.~\cite{ciuchi} (using Einstein phonons) into our 
multi-orbital DMFT.  The local intra- and inter-orbital Coulomb correlations 
are treated upto second order selfconsistent multiband IPT as usual. 
Since both Hubbard and electron-phonon interaction terms are local, their 
combined effect can be treated simultaneously within DMFT.
Contribution of the electron-phonon coupling to electronic
self energy is given by

$$ g^2\sum_{i\omega_n}G_{0}(p)D_{0}(\omega)=g^2[\frac{N_q+n_f(\zeta_p)}{ip_n+\omega_q-\zeta_p}+\frac{N_q+1-n_f(\zeta_p)}{ip_n-\omega_q-\zeta_p}] 
\eqno(5.2.1)$$ 

\noindent Where $N_q = \frac{1}{e^{\beta\omega_q}-1}$.\\
The full Hamiltonian, including excitonic coupling of phonons
is
$$H=\sum_{k,a,b,\sigma}(t_{k}^{ab}+\epsilon_a\delta_{ab})c^{\dagger}_{ka\sigma}c_{kb\sigma}
+U\sum_{i,\mu=a,b}n_{i\mu\sigma}
n_{i\mu-\sigma}+U_{ab}\sum_{i}n_{ia}n_{ib}+g\sum_{i}(c^{\dagger}_{a\sigma}c_{b\sigma}+h.c.)
(A^{\dagger}_{i}+A_{i})$$  
$$ -V\sum_{i}c^{\dagger}_{b\sigma}c_{b\sigma}
(1-c^{\dagger}_{a\sigma}c_{a\sigma})+\omega_0\sum_{i}A^{\dagger}_{i}A_{i}. \eqno(5.2.2)$$\\

We start with an initial ansatz for the self energy $\Sigma_{int}(\omega)=Un+
{A\Sigma_0^{(2)}(\omega)}$ where $\Sigma_0^{(2)}(\omega)$ is the second order contribution 
of electron-electron and interband excitons coupling to $A_{1g}$ phonons. The
IPT scheme is that we calculate lattice Green's function (G$_{fa}$) from this full
self energy, from which the bare Green's function is found via the Dyson equation: 
G$_{0a}^{-1} = G_{fa}^{-1}+\Sigma_{a}$.  Plugging this G$_{0a}$ back into the IPT, we 
obtain a new estimate of $\Sigma_{0a}^{(2)}$ and this procedure is iterated to convergence.

Having both Hubbard-like and el-ph couplings changes the estimate of $A_{ab}$ 
used in the interpolative self-energy in IPT as follows.
Following usual procedure, $A_{ab}$ is calculated from the condition
that it reproduce the leading behavior of the (of the exact atomic limit) 
self-energy at high frequency.  The leading behavior for large $\omega$ can be 
obtained by expanding the Green function in a continuous 
fraction \cite{gordon}:  $G_{fa}(k,\omega)=1/(\omega-\epsilon_{fa}-M_{1a}-
\frac{M_{2a}-M_{1a}^2}{\omega+\ldots})$.  
Here, $M_i$ denotes the $i$ th order moment of the density
 of states.  One can compute these quantities by evaluating a 
commutator\cite{nolting}, and for the model above,
 $M_{2a}-M_{1a}^2 = U_{ab}^2(n_{fa}(1-2n_{fa})+<n_{fa}n_{fb}>)+g^2(n_{fa+}+
n_{fa-})$.  Here, $n_{fa}$ is the number density calculated from the 
full Green's Function, $n_{fa+}$= $g^2\sum_\omega(G_f(\omega+\omega_q)
(N_q+n_f(\omega))$ and $n_{fa-}$= $g^2\sum_\omega(G_f(\omega-\omega_q)
(N_q+n_f(-\omega)).$ 
From the large frequency limit of (1),
 $\Sigma_{0a}^2(\omega)={U_{ab}^2n_{0a}(1-n_{0a})+
g^2(n_{0a-}+n_{0a+})}$.  Here, $n_{0a}$ is a fictitious number density 
of the `bare' Green's function.
Explicitly, $n_{0a+}$= $g^2\sum_\omega(G_0(\omega+\omega_q)
(N_q+n_f(\omega))$ and $n_{0a-}$= $g^2\sum_\omega(G_0(\omega-\omega_q)
(N_q+n_f(-\omega))$. Comparing with the exact high-frequency
limit, we thus have
$A_{ab}=\frac{ U_{ab}^2(n_{fa}(1-2n_{fa})+<n_{fa}n_{fb}>)
+g^2(n_{fa+}+n_{fa-})}{U_{ab}^2n_{0a}(1-n_{0a})+
g^2(n_{0a-}+n_{0a+})} $.


Within the DMFT approximation, the phonon self-energy turns out to be
$\Pi(q,\omega)=\frac{g^2\chi_c(q,\omega)}{1+g^2\chi_c(q,\omega)D_0(q,\omega)}$
where $\chi_c(q,\omega)$ is the usual charge-charge response
function~\cite{gunnar} and D$_0(q,\omega)$ is the bare phonon
Green's function.  $\chi_c(q,\omega)$ is estimated by a renormalised bubble 
contribution of the DMFT Green functions.  We ignore the irreducible vertex
corrections, since their contribution should be small for the small number of 
electrons and holes that characterise the two-band system close to an excitonic
insulator/liquid regime, and is a further approximation.

\end{document}